\documentclass[a4paper,12pt]{article}
\usepackage{amssymb, amsmath}
\usepackage[english]{babel}
\usepackage[applemac]{inputenc}
\usepackage[dvips]{graphicx}
\usepackage{amsfonts}
\usepackage{mathrsfs}
\usepackage{color}
\newcommand{\ud}{\mathrm{d}}
\newcommand{\shalf}{{\hbox{\small{$\frac{1}{2}\hskip 0.01cm$}}}}
\newtheorem{theorem}{Theorem}[section]
\newtheorem{proposition}{Proposition}[section]
\newtheorem{lemma}{Lemma}[section]

\newtheorem{definition}{Definition}[section]

\begin{document}


\begin{center}

{\Large{\textbf{On the nature of Bose-Einstein condensation\\
enhanced by localization}}}

\vspace{1cm}

\textbf{Thomas Jaeck\footnote{PhD student at UCD and Universit\'{e} de la M\'{e}diterran\'{e}e
(Aix-Marseille II, France), \\
e-mail: Thomas.Jaeck@ucdconnect.ie, phone: +353 1 7162571 },
Joseph V. Pul\'{e}\footnote{e-mail: joe.pule@ucd.ie, phone: +353 1 7162568}}\\
School of Mathematical Sciences, University College Dublin\\
Belfield, Dublin 4, Ireland

\vspace{1cm}

\textbf{Valentin A. Zagrebnov\footnote{e-mail: Valentin.Zagrebnov@cpt.univ-mrs.fr, phone: +33 491 26 95 04}}\\
Universit\'{e} de la M\'{e}diterran\'{e}e (Aix-Marseille II), \\ Centre de Physique Th\'{e}orique - UMR 6207,
Luminy - Case 907 \\ 13288 Marseille, Cedex 09, France

\vspace{1cm}

\begin{abstract}
\noindent
In a previous paper we established that for the perfect Bose gas
and the mean-field Bose gas with an external random or weak
potential, whenever there is generalized Bose-Einstein condensation in the eigenstates of the
single particle Hamiltonian, there is also
generalized condensation in the kinetic energy states. In these cases Bose-Einstein condensation
is produced or enhanced by the external potential.
In the present paper we establish a criterion for the absence of condensation in single
kinetic energy states  and prove that this criterion is satisfied for a class of random potentials
and weak potentials.
This means that the condensate is spread over an infinite number of states with low kinetic energy
without any of them being macroscopically occupied.
\end{abstract}

\end{center}


\vspace{0.5cm}

\textbf{AMS:} 82B10, 82B23, 81V70

\newpage
\section{Introduction}
\setcounter{equation}{0}
\renewcommand{\theequation}{\arabic{section}.\arabic{equation}}
It can be easily seen from the explicit formula for  the occupation numbers in the non-interacting
(\emph{perfect}) Bose gas that,
the condition for Bose-Einstein Condensation (BEC) to occur, is that the density of states of the
one particle Schr\"{o}dinger operator
decreases fast enough near the bottom of the spectrum. In the absence of any external potential, it
is known that this happens only in three dimension or higher. This is still true if one introduces a
mean-field interaction between particles. It has been known for some time that the
behavior of the density of states can be altered by the addition of suitable external potentials,
in particular \emph{weak}
potentials or \emph{random} potentials. The subject of this paper is the study of
models of the Bose gas in the presence of such external potentials.
The first case has been extensively studied, see e.g. \cite{P, VdBL}, where sufficient conditions on the external
potential were derived for the occurrence of BEC. In the random case, it has been
shown in \cite{LPZ} that the so-called \emph{Lifshitz tails}, which are a general feature of
disordered systems, see for instance \cite{PF}, are able to produce BEC. In both cases,
it is possible to obtain condensation even in dimensions $1$ or $2$.\\
While BEC has historically been associated with the macroscopic occupation of the ground state \emph{only},
it was pointed out in \cite{G} that this phenomena is more thermodynamically stable if
it is interpreted as as the macroscopic
accumulation of particles into an arbitrarily narrow band of energy above the ground
state, or \emph{generalized} BEC. While it is clear that condensation in the ground state implies generalized
BEC, there exist many situations in which the converse is not true. For instance,
it was shown in \cite{VdBL} that in the case of the weak potential, the condensate can be in one state, in
infinitely many states or even not in any state at all, depending on the external potential. These situations
correspond respectively to type I, II, III generalized BEC in the classification established by
{the Van den Berg, Lewis and Pul\'{e}}, see e.g. \cite{VdB-Lew-Pul}. In the random case, far less is known.
The only case for which a rigorous
proof of the exact type of BEC has been established is the Luttinger-Sy model, see \cite{LZ}, where it was shown
that the ground-state only is macroscopically occupied. As far as we know, for more complicated systems this is still an open question. The difficulty lies in the fact that the characterization of the distribution
of the condensate in individual states requires much more detailed knowledge about the spectrum than
the occurrence of generalized condensate. Indeed, for the latter, it is enough to know the asymptotic behavior
of the density of states, while for the former, one needs in addition to know how fast the gap between two
eigenvalues vanishes in limit.

In the physics literature the phenomenon of BEC is generally understood to be the macroscopic occupation
of the lowest kinetic energy (momentum) state, commonly referred as zero-mode condensation. We refer the reader to
\cite{JPZ} for a discussion of the motivation for this type of condensation.
This leads naturally to two questions in the case condensation is produced or enhanced by the addition of
external potentials.\\
The first one comes from the fact the condensates referred to here are to be found in the eigenstates of
the one-particle Schr\"{o}dinger operator and not the kinetic energy (momentum) states. Therefore, it is not
immediately clear, and in fact counter intuitive in the random case because of the lack of translation invariance,
that condensation occurs in the kinetic energy states as well. This problem has been
addressed in a previous paper, see \cite{JPZ}, where we have shown under fairly general assumptions
on the external potential (random or weak) that the amount of generalized BEC in the
eigenstates in turn creates a generalized condensate in the kinetic states, and moreover in the perfect gas the densities
of condensed particles are identical. These results were proved for the perfect Bose gas, and can be partially
extended to the mean-field Bose gas. Hence, the (generalized) condensation produced in these models
by the localization property of the one-particle Schr\"{o}dinger operator can be correctly described as
of \lq\lq Bose-Einstein'' type in the traditional sense. This opens up the possibility of
formulating a generalized version of the $c$-number Bogoliubov approximation (\cite{JPZ}, \cite{JZ}).
In the case of the weak external potential, perhaps this result is not so surprising since
the model is asymptotically translation invariant,
but in the random case, it is less obvious since the system
is translation invariant only in the sense that translates of the potential are equally probable and
therefore for a given configuration the system is not translation invariant.

Having established generalized BEC in the kinetic states, the next question is about the
fine structure of that condensate. In our paper \cite{JPZ}, we conjectured that the kinetic generalized BEC
is of type III, that is, no single kinetic state is macroscopically occupied, even though the amount
of generalized condensation is non-zero. Our motivation came from the fact that the fast decrease of the density
of states is usually associated with the corresponding eigenstates becoming \emph{localized} in the
infinite volume limit.
Hence, since the kinetic states (plane waves) and the (localized) general eigenstates are
\lq\lq asymptotically orthogonal'', it should follow that no condensation in any single-mode
kinetic energy state could occur,
independently of whether the (localized) ground state is macroscopically occupied or not.
In \cite{JPZ}, we were
able to prove this conjecture in a simple example, the Luttinger-Sy model.
Our proof in that case used the absence of
tunneling effect specific to that model, which we can interpret as \lq\lq perfect localization".

In this paper we give a proof of the conjecture under a fairly weak localization hypothesis
and then we consider a family of continuous
random models and a general class of weak external potential for which we
are able to establish this localization property. Our results hold for
both the perfect and mean-field Bose gas, and for any dimension. Note that, in addition to clarifying the nature
of these condensates in low dimensions, we obtain an {\emph{unexpected conclusion}}. Indeed,
we show that the presence of randomness or a weak potential, however small, prevents condensation
from occurring in any kinetic state, even if the corresponding \emph{free} Bose gas (without external potential)
exhibits zero-mode condensation (isotropic system in dimension 3, for example). This
emphasizes the importance of the concept of generalized BEC.

The structure of the paper is as follows: in Section \ref{Notation and models} we give the
general setting for which our results are applicable
and discuss generalized condensation in the kinetic energy states,
while in Section \ref{Localization and kinetic single-state BEC}
we derive a criterion for the absence of condensation
into any single kinetic energy state. In Section \ref{Proof of the localization condition}
we establish that this criterion is
satisfied for a class of random potentials (Subsection \ref{Random potentials})
and for weak (scaled) potentials (Subsection \ref{Weak external potentials}).

\section{Notation and models}\label{Notation and models}
\setcounter{equation}{0}
\renewcommand{\theequation}{\arabic{section}.\arabic{equation}}

Let $\{\Lambda_{l}:= (-l/2, l/2)^{d}\}_{l \geqslant 1}$ be a sequence of hypercubes of side $l$ in
$\mathbb{R}^{d}$, centered at the origin of coordinates with volumes $V_{l} = l^{d}$.
We consider a system of identical bosons, of mass $m$, contained
in $\Lambda_{l}$. For simplicity, we use a system of units such that $\hbar = m = 1$. First we define the
self-adjoint one-particle kinetic-energy operator of our system by:
\begin{eqnarray}\label{Kinetic-energy-operator}
h_{l}^{0} := -\shalf \Delta_D,
\end{eqnarray}
acting in the Hilbert space $\mathscr{H}_{l} := L^{2} (\Lambda_{l})$, where $\Delta$ is the usual Laplacian.
The subscript $D$ stands for \textit{Dirichlet} boundary conditions. We denote
by $\{\psi_{k}^{l}, \varepsilon_{k}^{\l}\}_{k \geqslant 1}$ the set of normalized
eigenfunctions and eigenvalues corresponding to $h_{l}^{0}$. By convention, we order the eigenvalues
(counting multiplicity) as $0 < \varepsilon_{1}^{l} \leqslant  \varepsilon_{2}^{l} \leqslant
\varepsilon_{3}^{l}\dots\,\,$. Note that all kinetic states satisfy the following bound
\begin{eqnarray}\label{upper-bound-kinetic-eigenstates-modulus}
\vert \psi_{k}^{l} (x) \vert \, \leqslant \, l^{-d/2}\, .
\end{eqnarray}
Next we define the Hamiltonian with an external potential
\begin{eqnarray}\label{Schrodinger-operator}
h_{l} := h_{l}^{0} + \, v_l,
\end{eqnarray}
also acting in $ \mathscr{H}_l$, where $v_l:\Lambda_l \mapsto [0,\infty)$ is positive and bounded.
Let $\{\phi_{i}^{l}\}_{i \geqslant 1}$ and $\{E_i^l\}_{i \geqslant 1}$ be respectively the
sets of normalized eigenfunctions and corresponding eigenvalues of $h_{l}$. Again, we order the eigenvalues
(counting multiplicity) so that $E_{1}^{l} \leqslant E_{2}^{l} \leqslant E_{3}^{l}\dots \,\,$.
Note that the \textit{non-negativity} of the potential implies {that} $E_{1}^{l} > 0$.
We shall also assume that the lower end of the spectrums of $h_{l}^{0}$ and $h_{l}$ coincide in the limit
$l \rightarrow \infty$, that is, $\lim_{l \rightarrow \infty} E_{1}^{l} = 0$.
This assumption will be proved for the models considered in this paper.

Now, we turn to the many-body problem. Let $\mathscr{F}_{l}:= \mathscr{F}_{l}(\mathscr{H}_{l})$ be the
symmetric Fock space constructed over $ \mathscr{H}_{l}$. Then $H_{l}:={\rm{d\Gamma}}(h_{l})$ denotes
the second quantization of the \textit{one-particle} Schr\"{o}dinger operator $h_{l}$ in $\mathscr{F}_{l}$.
Note that the operator $H_{l}$ acting in $\mathscr{F}_{l}$ has the form:
\begin{equation}\label{Fock-one-part}
H_{l} =  \sum_{i \geqslant 1}  E_i^l \   a^{*}(\phi_{i}^{l}) a(\phi_{i}^{l}),
\end{equation}
where $a^{*}(\varphi),  a(\varphi)$ are the creation and annihilation operators (satisfying the boson
\textit{Canonical Commutation Relations}) for the one-particle state $\varphi \in \mathscr{H}_l$. Then, the
grand-canonical Hamiltonian of the perfect Bose gas in an external potential is given by:
\begin{eqnarray}\label{perfect-multi.part.Hamiltonian}
H_{l} - \mu N_{l} \, = \, \sum_{i \geqslant 1}  (E_i^l - \mu)\   N_{l}
(\phi_{i}^{l})
\end{eqnarray}
where $N_{l}(\phi) := a^{*}(\phi) a(\phi)$ is the operator for the number of particles
in the normalized state $\phi$, $N_{l} := \sum_{i} N_{l}(\phi_{i}^{l})$  is the operator for the total number of
particles in $\Lambda_l$ and $\mu$ is the chemical potential.

The results in this paper hold for the mean-field Bose gas whose many particle Hamiltonian, $H_{l} (\mu)$, is obtained by adding a mean-field term to (\ref{perfect-multi.part.Hamiltonian}): 
\begin{eqnarray}\label{mean-field-many-part-hamiltonian}
H_{l} (\mu) \, := \, H_{l} - \mu N_{l} \, + \, \frac{\lambda}{2V_{l}} \, N_{l}^{2},
\end{eqnarray}
where $\lambda$ is a non-negative parameter. Of course the results are valid also for the perfect Bose gas ($\lambda=0$). 

We recall that the thermodynamic equilibrium Gibbs state $\langle - \rangle_{l}$ associated with the
Hamiltonian $H_{l}(\mu)$ is defined by:
\begin{eqnarray*}
\langle A\rangle_{l} :=
\frac{\textrm{Tr}_{ \mathscr{F}_{l}}\{\exp ( -\beta H_{l}(\mu_l) ) A\}}{\textrm{Tr}_{ \mathscr{F}_{l}}
\exp ( -\beta H_{l}(\mu_l) )}
\end{eqnarray*}
where the value of $\mu$, $\mu_l$ is determined by fixing the mean density $\overline{\rho} >0$:
\begin{eqnarray}\label{constraint}
\frac{1}{V_{l}} \langle N_{l}\rangle_{l} =\overline{\rho}.
\end{eqnarray}
When referring specifically to the perfect Bose gas state, we shall use the notation $\langle - \rangle_{l}^{0}$.
For simplicity, in the sequel we shall omit the explicit mention of the dependence on the
thermodynamic parameters $(\beta,\mu)$ unless it is necessary to refer to them.

A normalized single particle state $\varphi$ is macroscopically occupied if
\begin{eqnarray*}
\lim_{l \rightarrow \infty} \, \frac{1}{V_{l}} \langle N_{l}(\varphi)\rangle_{l} >0
\end{eqnarray*}
and in particular there is condensation in the ground state if
\begin{eqnarray*}
\lim_{l \rightarrow \infty} \, \frac{1}{V_{l}} \langle N_{l}(\phi_1^l)\rangle_{l} >0.
\end{eqnarray*}
The concept of generalized condensation
consists in considering the possible macroscopic occupation of an arbitrary small band of energies
at the bottom of the spectrum.
To be more precise, we say that there is generalized condensation in the states $\phi_{i}^{l}$ if
\begin{eqnarray*}
\lim_{\delta \downarrow 0} \, \lim_{l \rightarrow \infty} \, \frac{1}{V_{l}} \, \sum_{i: E^l_{i} \leqslant \delta}
\langle N_{l}(\phi_{i}^{l})\rangle_{l}>0.
\end{eqnarray*}
It is clear that the usual one-mode condensation implies generalized condensation, however the
converse is not true. Indeed, as was first established by the Dublin School in the eighties
\cite{VdB-Lew-Pul}, it is possible to classify
generalized condensation into three types. Type I condensation, when a finite number of states are macroscopically
occupied (which includes the most commonly known notion of BEC as condensation in the ground state only),
type II condensation,
when condensation occurs in an infinite number of states, and finally type III, when, although the amount of
generalized condensation is non-zero, \emph{no} individual state is macroscopically occupied.
One can easily show that in the perfect Bose gas,
under fairly general assumptions, for both random and weak positive potentials,
there is indeed generalized condensation in a suitable range of density (or temperature).

In \cite{JPZ} we discussed the possibility
of generalized condensation not in the states $\phi_{i}^{l}$ but in the kinetic energy states $\psi_{k}^{l}$.
For both random and weak positive potentials, we established that for models which are diagonal
in the occupation numbers of the eigenstates of the Hamiltonian (\ref{Schrodinger-operator}),
the density of \emph{generalized BEC} in the kinetic states is never less then that
in the eigenstates of the single particle Hamiltonian.
To be more precise, we proved that
\begin{eqnarray*}
\lim_{\delta \downarrow 0} \, \lim_{l \rightarrow \infty} \, \frac{1}{V_{l}} \, \sum_{k: \varepsilon^l_{k} < \delta}
\langle N_{l}(\psi_{k}^{l})\rangle_{l}
\geqslant
\lim_{\delta \downarrow 0} \, \lim_{l \rightarrow \infty} \, \frac{1}{V_{l}} \, \sum_{i: E^l_{i} < \delta}
\langle N_{l}(\phi_{i}^{l})\rangle_{l}.
\end{eqnarray*}
We also showed that in the case of the perfect gas, the two quantities in the above inequality are equal.
Here, we shall give a \lq\lq localization criterion" on the states $\phi_{i}^{l}$ so that the condensation
in the kinetic energy states
$\psi_{k}^{l}$ is of type III,
that is no kinetic energy state is macroscopically occupied.

It is easy to see that the mean-field gas satisfies the following commutation relation
\begin{eqnarray}\label{general-eigenstate-invariance}
[H_{l}(\mu), N_{l}(\phi_{j}^{l})] = 0, \quad \textrm{for all}\  j.
\end{eqnarray}
This property implies that $\langle a^{*}(\phi_{i}^{l}) a(\phi_{j}^{l})\rangle_{l}=0$ if $i \neq j$ and allows
us to obtain a simple relation between the mean occupation numbers for the $\psi_{k}^{l}$'s and $\phi_{k}^{l}$'s:
\begin{eqnarray}\label{expansion-occupation-number-kinetic-state-to-random-state}
\frac{1}{V_{l}} \langle N_{l}(\psi_{k}^{l})\rangle_{l} = \frac{1}{V_{l}} \langle a^{*}(\psi_{k}^{l})
a(\psi_{k}^{l})\rangle_{l} \, &=& \, \frac{1}{V_{l}} \,\sum_{i,j} \, (\phi_{i}^{l},\psi_{k}^{l})
\overline{(\phi_{j}^{l},\psi_{k}^{l})} \,
\langle a^{*}(\phi_{i}^{l}) a(\phi_{j}^{l})\rangle_{l} \\
&=&  \frac{1}{V_{l}} \,\sum_{i}  \,  |(\phi_{i}^{l},\psi_{k}^{l})|^{2} \,
\langle N_{l}(\phi_{i}^{l})\rangle_{l}.\nonumber
\end{eqnarray}
Finally, we want to point out that it may be possible to extend the results of this paper to a more general
class of interacting
Bose gases. More precisely, consider a class of \lq\lq diagonal'' interactions defined by
\begin{eqnarray*}
U_{l} \, := \, \frac{\lambda}{V_{l}} \sum_{i,j} \, a_{i,j} \, N_{l}(\phi_{i}) N_{l}(\phi_{j})
\end{eqnarray*}
with suitable assumptions on the coefficients $a_{i,j}$ in order to make the associated many-particle Hamiltonian
well-defined, that is self-adjoint and bounded below. Note that the mean-field gas
(\ref{mean-field-many-part-hamiltonian}) is a particular case of this class,
in which $a_{i,j} = \delta_{i,j}$ (with a shift in the chemical potential). It is easy to see that the condition
(\ref{general-eigenstate-invariance}) is satisfied. However we shall also need the monotonicity of
the mean occupation numbers $\langle N_{l}(\phi_{i}^{l})\rangle_{l}$ (see Lemma
\ref{lemma-monotonicity-perfect-and-meanfield-gas}), which
so far we are unable to prove beyond the mean-field case.

In the next section we use the expansion (\ref{expansion-occupation-number-kinetic-state-to-random-state}) to obtain
a localization criterion for the absence of single mode condensation in the kinetic energy states.


\section{Localization and kinetic single-state BEC}\label{Localization and kinetic single-state BEC}
\setcounter{equation}{0}
\renewcommand{\theequation}{\arabic{section}.\arabic{equation}}
First we shall prove the following lemma which is trivial for the perfect gas. For the mean-field Bose gas it was
proved by Fannes and Verbeure, \cite{FV2}, using correlations inequalities.
Here we present  an alternative proof, based only on a convexity argument.
\begin{lemma}\label{lemma-monotonicity-perfect-and-meanfield-gas}
For the mean-field Bose gas, i.e. for a bosonic system with Hamiltonian
(\ref{mean-field-many-part-hamiltonian}), the function $i \rightarrow \langle N_{l}(\phi_{i})\rangle_{l}$ is non-increasing.
\end{lemma}
\textbf{Proof:}

\noindent
Let us define $f:\mathbb{R}_+\mapsto \mathbb{R}$ by
\begin{eqnarray*}
f(t) &:=& \beta^{-1} \, \ln \, \textrm{Tr} \, \mathrm{e}^{- \beta H_{l}(\mu;t)},\\
\textrm{where} \quad H_l(\mu;t) &:=& H_{l}(\mu) + t (N_{l}(\phi_{m}^{l}) - N_{l}(\phi_{n}^{l})),
\end{eqnarray*}
for some $1 \leqslant m < n$. It follows that
\begin{eqnarray*}
f'(0) \, =  \, \langle N_{l}(\phi_{n}^{l}) - N_{l}(\phi_{m}^{l})\rangle_{l}
\end{eqnarray*}
and since the function $f$ is convex, we have
the following inequality
\begin{eqnarray}\label{monotonicity-occupation-number-proof-meanfield}
\langle N_{l}(\phi_{n}^{l}) - N_{l}(\phi_{m}^{l})\rangle_{l} \, \leqslant \, f'(t),
\end{eqnarray}
for any $t \geqslant 0$. Now we set $t = \frac{1}{2} (E_{n}^{l} - E_{m}^{l})$. Note that with this choice
$t\geqslant 0$, since we have assumed that $m < n$. From the explicit
expression (\ref{mean-field-many-part-hamiltonian}) for $H_{l}(\mu)$, we have
\begin{eqnarray*}
H_{l}(\mu;t) \, = \, \sum_{i \neq m,n} (E_{i}^{l} - \mu) N_{l}(\phi_{i}^{l}) \,
+ \, \frac{\lambda}{2V_{l}} \, N_{l}^{2}\, + \,
(\frac{E_{m}^{l} + E_{n}^{l}}{2} - \mu) N_{l}(\phi_{m}^{l}) \, + \, (\frac{E_{m}^{l} + E_{n}^{l}}{2}
- \mu) N_{l}(\phi_{n}^{l}).
\end{eqnarray*}
Since the mean-field term in (\ref{mean-field-many-part-hamiltonian}) is symmetric with respect
to a permutation of any two eigenstate
indices $i,j$, it follows that $H_{l}(\mu;t)$ is symmetric with respect to the exchange of $m$ and $n$. Hence
\begin{eqnarray*}
f'(t)  \, = \, \frac{\textrm{Tr} \,\big(N_{l}(\phi_{n}^{l}) - N_{l}(\phi_{m}^{l})\big)
\,  \mathrm{e}^{- \beta H_{l}(\mu;t)}}{\textrm{Tr} \,  \mathrm{e}^{- \beta H_{l}(\mu;t)}} \, = \, 0,
\end{eqnarray*}
which in view of (\ref{monotonicity-occupation-number-proof-meanfield}) gives
\begin{eqnarray*}
\langle N_{l}(\phi_{n}^{l}) - N_{l}(\phi_{m}^{l})\rangle_{l} \, \leqslant \, 0,
\end{eqnarray*}
and the lemma follows since $m < n$ are arbitrary. \hfill $\square$

Let us introduce the notation
\begin{eqnarray*}
\rho_{i}^{l} \, := \, \frac{1}{V_{l}} \,  \langle N_{l}(\phi_{i}^{l})\rangle_{l}.
\end{eqnarray*}
With this notation we can write the standard fixed density condition (\ref{constraint}) for a given density
$\overline{\rho}$, as
\begin{eqnarray*}
\sum_{i} \, \rho_{i}^{l} \, = \, \overline{\rho},
\end{eqnarray*}
and so for any $N\in \mathbb{N}$,
\begin{eqnarray*}
\sum_{i=1}^{N} \, \rho_{i}^{l} \, \leqslant \, \overline{\rho}.
\end{eqnarray*}
Letting
\begin{eqnarray*}
\rho_{i} \, := \, \limsup_{l \rightarrow \infty} \, \rho_{i}^{l},
\end{eqnarray*}
and taking the infinite volume limit, we then get
\begin{eqnarray*}
\sum_{i=1}^{N} \, \rho_{i} \, = \, \limsup_{l \rightarrow \infty} \, \sum_{i=1}^{N} \,
\rho_{i}^{l} \, \leqslant \, \overline{\rho}.
\end{eqnarray*}
Letting $N$ tend to infinity, this gives $\sum_{i=1}^{\infty} \rho_{i} \leqslant \overline{\rho}$, and hence,
for any $\varepsilon > 0$,
there exists $i_{0} < \infty$ such that $\rho_{i_{0}} \, < \, \varepsilon$. Splitting up the sum
in (\ref{expansion-occupation-number-kinetic-state-to-random-state}) and using the
monotonicity property (see Lemma \ref{lemma-monotonicity-perfect-and-meanfield-gas}), property
(\ref{upper-bound-kinetic-eigenstates-modulus})
and the fact that the kinetic eigenfunctions $\psi^l_{k}$ are normalized, we obtain
\begin{eqnarray*}
\frac{1}{V_{l}} \, \langle N_{l}(\psi_{k}^{l})\rangle_{l} \, &=& \, \sum_{i \leqslant i_{0}}  \,
|(\phi_{i}^{l},\psi_{k}^{l})|^{2} \,
\rho_{i}^{l} \, + \, \sum_{i > i_{0}}  \,  |(\phi_{i}^{l},\psi_{k}^{l})|^{2} \,
\rho_{i}^{l} \\
&\leqslant& \sum_{i \leqslant i_{0}}  \,  |(\phi_{i}^{l},\psi_{k}^{l})|^{2} \,
\rho_{i}^{l} \, + \,  \rho_{i_{0}}^{l} \sum_{i> i_{0}}  \,  |(\phi_{i}^{l},\psi_{k}^{l})|^{2}\\
&\leqslant& \overline{\rho} \, \sum_{i \leqslant i_{0}}  \,  |(\phi_{i}^{l},\psi_{k}^{l})|^{2} \,
+ \, \rho_{i_{0}}^{l}\\
&\leqslant&  \overline{\rho} \, \sum_{i \leqslant i_{0}} \,  \big( l^{-d/2} \, \vert\vert \phi_{i}^{l} \vert\vert_{1} \big)^{2} \,
+ \,  \rho_{i_{0}}^{l}\, .
\end{eqnarray*}
Therefore if $l^{-d/2} \, \vert\vert \phi_{i}^{l} \vert\vert_{1}\to 0$ as $l\to \infty$ for each $i$, then
\begin{eqnarray*}
\limsup_{l \rightarrow \infty} \, \frac{1}{V_{l}} \, \langle N_{l}(\psi_{k}^{l})\rangle_{l} \, \leqslant \,
\varepsilon,
\end{eqnarray*}
and since $\varepsilon$ is arbitrary
\begin{eqnarray*}
\lim_{l \rightarrow \infty} \, \frac{1}{V_{l}} \, \langle N_{l}(\psi_{k}^{l})\rangle_{l} =0.
\end{eqnarray*}

The above argument leads us to define the following localization criterion for the absence of
single mode condensation in the kinetic energy states.
\begin{definition}\label{localization-property-definition}
We call an eigenfunction $\phi_{i}^{l}$ localized if it satisfies the following condition
\begin{eqnarray}\label{general-localization-definition}
\lim_{l \rightarrow \infty} \, \frac{1}{l^{d/2}} \, \int_{\Lambda_{l}} \ud x \, \vert \phi_{i}^{l}(x) \vert \,\,
= \,\, 0.
\end{eqnarray}
\end{definition}
{Note that this localization condition is not as strong as the usual localization property, in the following sense.
While, localization is
frequently understood to be associated with the persistence of a pure point spectrum in the limit
$l \rightarrow \infty$, at least near the bottom of the spectrum, the presence of
a pure point spectrum is not necessary for
the condition (\ref{general-localization-definition}) to hold for all eigenfunctions. Indeed it may happen
that (\ref{general-localization-definition}) is satisfied and the infinite volume Schr\"{o}dinger operator
has only absolutely continuous spectrum.}

In \cite{JPZ} we conjectured that the kinetic generalized BEC observed in the random models is in fact
of type III, and gave a proof in a simple case, the Luttinger-Sy model. In the above argument we proved that
our conjecture is correct under the fairly weak localization hypothesis (\ref{general-localization-definition}).
We formulate this result in the following theorem.
\begin{theorem}\label{type3-BEC-random-systems}
Assume that the eigenfunctions $\phi_{i}^{l}$ are localized in the sense of
(\ref{general-localization-definition}) for all $i$. Then, for the mean-field Bose gas, no kinetic state $\psi_{k}^{l}$ can be macroscopically
occupied, that is
\begin{eqnarray}
\lim_{l \rightarrow \infty} \, \frac{1}{V_{l}} \, \langle N_{l}(\psi_{k}^{l})\rangle_{l} \,\, = \,\, 0,
\end{eqnarray}
which implies in particular that any possible kinetic generalized BEC in these models is of type III.
\end{theorem}
In this paper, we provide two classes of externals potential for which we can prove localization in the sense of
(\ref{general-localization-definition}). The first one is a class of \textit{random} external potentials, the second
involves \textit{weak} external potentials.
\section{Proof of the localization condition}\label{Proof of the localization condition}
\subsection{Random potentials}\label{Random potentials}
\setcounter{equation}{0}
\renewcommand{\theequation}{\arabic{section}.\arabic{equation}}

Before we specify the random model under consideration, let us emphasize again that the localization
property (\ref{general-localization-definition})
is very different from what is usually called \lq\lq exponential localization'' in the literature about random
Schr\"{o}dinger operators (see for example \cite{S}).
In the standard literature localization refers to the eigenfunctions of the infinite volume Hamiltonian and
requires these functions,
with energies in some band, to decay very fast, in many cases
exponentially. This implies that the spectrum is pure point in that band. In our case we are dealing with
eigenfunctions
in finite volume with energies tending to zero as the volume increases and so these bear no relation to the
infinite volume eigenfunctions.
In particular, our localization condition (\ref{general-localization-definition}) does \emph{not} imply that
the spectrum is discrete in the thermodynamic limit.
While we only need the $L^1$ norm not to diverge too fast, because our eigenfunctions depend crucially on the
volume and in particular,
because we do not work at a fixed energy but with volume dependent eigenvalues, we have to deal with the
additional problem of
controlling the finite-volume behavior.
However, we find that in fact the \emph{multiscale analysis} developed for the infinite volume case can be
adapted to establish our localization condition.

The model studied in this section is taken from \cite{S}. It consists of impurities located at points of the
lattice $\mathbb{Z}^{d}$,
with appropriate assumptions over the single-impurity potential, mainly designed to obtain independence between
regions which are sufficiently far away from each other.
Let us make it more explicit by giving some definitions. In the rest of this section, we shall
denote by $\Lambda_{l}(x)$ the cubic box of side $l$ centered at $x$.
The single-site potential $f$, $\Lambda_{1}(0) \rightarrow \mathbb{R}$ has the following properties:
\begin{enumerate}
  \item $f$ is bounded;
  \item there is $\sigma > 0$ such that $f(x) \geqslant \sigma$ for all $x \in \Lambda_{1}(0)$.
\end{enumerate}
The randomness in this model is given by varying the strength of each impurity. For this purpose, we define
a single-site (probability) measure $\mu$, with $\mathrm{supp}(\mu) = [0,a]$ for a finite $a$. We will assume
that $\mu$ is
H\"{o}lder-continuous, that is for some $\alpha > 0$,
\begin{eqnarray}\label{Holder-continuity-proba-distr-continuous-case}
\sup_{\{ s,t \}} \big\{ \mu([s,t]): 0 \leqslant t-s \leqslant \eta \big\} \, \leqslant \, \eta^{\alpha}, \quad
\forall \,\, 0 \leqslant \eta \leqslant 1
\end{eqnarray}
The random potential is then defined by
\begin{eqnarray}\label{random potential}
v^{\omega}(x) \, := \, \sum_{k \in \mathbb{Z}^{d}} \, q^{\omega} (k) \, f(x-k),
\end{eqnarray}
where the $q^{\omega}(k)$'s are i.i.d. random variables distributed according to $\mu$.
We denote by $(\Omega, \mathcal{F}, \mathbb{P})$ the associated probability space,
and by $\omega \in \Omega$ a particular realization of the
random potential. Note that by property 2 and the fact that $a<\infty$,
there exists a non-random $M < \infty$ such that $v^{\omega}(x) < M$ for any $x$ and all $\omega$.

The one-particle random Schr\"{o}dinger operator in finite volume is then given as in (\ref{Schrodinger-operator}) by
\begin{eqnarray}\label{Schrodinger-operator-random}
h_{l}^{\omega} \, = \, h_l^0+v_l^{\omega},
\end{eqnarray}
where $v_l^{\omega}$ is the restriction of $v^{\omega}$ to $\Lambda_l$.
The eigenfunctions and eigenvalues of $h_{l}^{\omega}$ are denoted by $\phi_i^{\omega,l}$ and $E_i^{\omega,l}$
respectively.
We denote by $h_{l}^{\omega}(x)$ the restriction of the Schr\"{o}dinger operator $-\shalf \Delta  +  v^{\omega}$
to the region $\Lambda_{l}(x)$, with Dirichlet boundary conditions.

Before we establish the localization criterion (\ref{general-localization-definition})
we prove our assumption
that the eigenvalues of $h_{l}^{\omega}$ tend to zero as $l$ tends to $\infty$:
\begin{lemma}\label{eigenvalues}
With probability one, for each $i$
\begin{eqnarray}
\lim_{l\to \infty}E_i^{\omega,l}=0.
\end{eqnarray}
\end{lemma}
\textbf{Proof:} \ Let $\nu$ denote the limiting integrated density of states for the Hamiltonians $h_l^{\omega}$,
that is
for any Borel subset $A\subset \mathbb{R}_+$,
\begin{eqnarray}
\nu(A):=\lim_{l\to \infty}\frac{1}{V_l}\sharp \{i\, :\, E_i^{\omega,l}\in A\}.
\end{eqnarray}
Since by ergodicity $\nu$ is nonrandom (see for example Theorem 5.18 in \cite{PF}),
it is clearly sufficient to prove that for every $E>0$, $\nu([0,E])>0$.
To do this we start from the following inequality (see equation (4) in \cite{KM}),
\begin{eqnarray}\label{DN-bracketting-for-IDS}
\nu ([0,E]) \, \geqslant \, \frac{1}{V_L} \, \mathbb{E} \big( \sharp \{i: E_{i}^{\omega,L} \leqslant E\}  \big)
\geqslant \, \frac{1}{V_L} \, \mathbb{P} \{\omega: \, E_{1}^{\omega,L} \, \leqslant \, E\}.
\end{eqnarray}
which is satisfied for any $L > 0$.
From the min-max principle, we obtain
\begin{eqnarray}\label{proof-Lifshitz-tails-Stollmann-min-max-principle}
E_{1}^{\omega,L} \, \leqslant \, \varepsilon_{1}^{L} \, + \, \int_{\Lambda_{L}} \ud x \,
\vert \psi_{1}^{L} \vert^{2} \, v^{\omega}(x)
\end{eqnarray}
where $\varepsilon_{1}^{L}$ is the first kinetic eigenvalue and $\psi_{1}^{L}$ the corresponding eigenfunction.
Since $\vert \psi_{1}^{L}(x) \vert^2 \leqslant 1/V_L$, we have
\begin{eqnarray}\label{bound3}
E_{1}^{\omega,L} \, \leqslant \, \varepsilon_{1}^{L} \, + \, \frac{1}{V_L}\int_{\Lambda_{L}} \ud x \, v^{\omega}(x)
\leqslant \, \varepsilon_{1}^{L} \, + \, \frac{A}{V_L}\sum_{k\in \mathbb{Z}^d\cap \Lambda_{L}} q^\omega(k).
\end{eqnarray}
where $A:= \, \int_{\Lambda_1} \ud x \, f(x)$.
Letting $L := \pi (E/2)^{-1/2}$ so that $\varepsilon_{1}^{L} = E/2$, (\ref{DN-bracketting-for-IDS}) and (\ref{bound3})
give
\begin{eqnarray}
\nu ([0,E]) \, \geqslant \,
\frac{1}{V_L} \, \mathbb{P} \{\omega: \, \sum_{k\in \mathbb{Z}^d\cap \Lambda_{L}} q^\omega(k) \leqslant \, EV_L/2A\}.
\end{eqnarray}
Since the right-hand side of the last inequality is strictly positive, the lemma is proved. \hfill $\square$

The rest of this subsection is devoted to proving that this model satisfies our localization assumption
(\ref{general-localization-definition}).
For this purpose we need a result from multiscale analysis which exists in various forms
in the literature (see references in \cite{S}). For convenience here we follow the version in \cite{S}.

Adhering to the terminology of \cite{S},  we first define so-called \lq\lq good boxes'':
\begin{definition}
Given $x \in \mathbb{Z}^{d}$, a scale $l \in 2\mathbb{N} + 1$, an energy $E$, a rate of decay $\gamma > 0$,
we call the box $\Lambda_{l}(x)$ $(\gamma, E)$-good for a particular realization $\omega$ of the
random potential (\ref{random potential}) if $E\notin \sigma(h_{l}^{\omega}(x))$ and
\begin{eqnarray}\label{exp-bound}
\vert\vert \chi_{l}^{out} \, (h_{l}^{\omega}(x) - E)^{-1} \, \chi_{l}^{int} \vert\vert \,\, \leqslant \,\, e^{-\gamma l}.
\end{eqnarray}
Here $\sigma(h_{l}^{\omega}(x))$ denotes the spectrum of $h_{l}^{\omega}(x)$,
the norm in (\ref{exp-bound}) refers to the
operator norm in $L^{2}(\Lambda_{l}(x))$, and $\chi_{l}^{int}, \chi_{l}^{out}$
are the characteristic functions of the regions $\Lambda_{l}^{int}(x), \Lambda_{l}^{out}(x)$ respectively,
which we define as follows
\begin{eqnarray*}
\Lambda_{l}^{int}(x) \, := \, \Lambda_{l/3}(x), \quad \Lambda_{l}^{out}(x) \, := \,
\Lambda_{l}(x) \setminus \Lambda_{l-2}(x).
\end{eqnarray*}
\end{definition}

Our proof depends crucially on the following important multiscale analysis result extracted from \cite{S}.
We refer the reader
to Theorem 3.2.2 and Corollary 3.2.6 for the general multiscale analysis argument and to Theorems 2.3.2, 2.2.3 and 2.4.1
for proving that this particular model satisfies the necessary conditions required for multiscale analysis.
\begin{proposition}\label{multiscale-continuous-case}
Assume that $h_{l}^{\omega}$ is as above with random potential given by
(\ref{random potential}).
Then for any $\zeta >0$ and any $\alpha \in \big(1,2-(4d/(4d+\zeta))\,\big ]$, there exist a sequence
$\{l_{k}\}, k\geqslant 1$, satisfying
$l_1\geqslant 2$ and $l_{k-1}^{\alpha} \leqslant l_{k} \leqslant l_{k-1}^{\alpha} + 6$ for $k\geqslant 2$,
and constants $r >0$ and $\gamma> 0$
such that if $I:=[0,r]$,
\begin{eqnarray}\label{multiscale-result-continuous-case}
\mathbb{P} \Big\{ \omega: \ \mathrm{for\ all}\ E \in I, \ \mathrm{either} \ \Lambda_{l_{k}}(x) \ \mathrm{or} \
\Lambda_{l_{k}}(y) \ \mathrm{is}\ (\gamma,E)\textrm{-} \mathrm{good} \, \Big\}
\, \geqslant \, 1\, - \, (l_{k})^{-2\zeta},
\end{eqnarray}
for all $k\geqslant1$ and for all $x,y  \in \mathbb{Z}^{d}$, satisfying $|x-y| > l_{k}$.

\end{proposition}

For our proof we need also the \textit{Eigenfunction
Decay Inequality}. We state it in a convenient form for our purpose, and refer the reader to \cite{S} (Lemma 3.3.2)
for a detailed proof. Note that this inequality has to be understood for a given realization $\omega$.
\begin{proposition}\label{eigenfunction-decay-inequality-continuous-case}
Let $h_{l}^{\omega}$ be defined as above, and $\phi_{i}^{\omega,l}$ to be one eigenfunction
with eigenvalue $E_{i}^{\omega,l}$.
Let $x \in \Lambda_{l}$, such that $\Lambda_{l_{k}}(x) \subset \Lambda_{l}$. If $E_{i}^{\omega,l}$ does
not belong to the spectrum of $h_{l_{k}}^{\omega}(x)$, then the following inequality holds
\begin{eqnarray}\label{eigenfunction-decay-inequality}
 \| \chi_{l_{k}}^{int}(x) \phi_{i}^{\omega,l} \| \, \, \leqslant \, \, \kappa \| \chi_{l_{k}}^{out}(x)
(h_{l_{k}}^{\omega}(x) - E_{i}^{\omega,l})^{-1} \chi_{l_{k}}^{int}(x) \|,
\end{eqnarray}
where the norms are $L^{2}(\Lambda_{l})$-norm, and $\kappa$ is a constant depending only on $M$.
\end{proposition}

We are now ready to prove that for our model the localization condition
(\ref{general-localization-definition}) is satisfied.

\begin{lemma}\label{localization-continuous-case}
Assume that $h_{l}^{\omega}$ is as in (\ref{Schrodinger-operator-random}) with random potential given by
(\ref{random potential}). Then almost surely, for all $i$,
\begin{eqnarray}\label{localization-condition-continuous-case}
\lim_{l \rightarrow \infty} \, \frac{1}{V_{l}^{1/2}} \, \int_{\Lambda_{l}} \, dx \,
\vert \phi_{i}^{\omega,l} (x) \vert \,\, = \,\, 0.
\end{eqnarray}
\end{lemma}
\textbf{Proof}:
We first choose $0 < \delta < 1/7$ and $\zeta>(2d+1)/2\delta$ and then we take the constants $\alpha$,
$\gamma$ and $r$ and the sequence
$\{l_{k}\}$ to be those obtained in Proposition \ref{multiscale-continuous-case} for this value of $\zeta$. For
a given scale $l$ large enough we pick $k = k(l)$ satisfying
\begin{eqnarray*}
\frac{1}{\ln \alpha} \ln \left( \frac{\delta \ln l}{ \ln l_{1}}\right) \, < \, k \, < \, \frac{1}{\ln \alpha}
\ln \left( \frac{(1 - \delta) \ln l}{ \ln (l_{1} + 6)}\right).
\end{eqnarray*}
The fact that $\delta <1/7$ ensures that there exists such an integer $k$.
Then, by Proposition \ref{multiscale-continuous-case}, we have
\begin{eqnarray}\label{allowed-interval-for-lk-continuous-case}
l^{\delta} < l_{k} < l^{1-\delta}.
\end{eqnarray}
Now let us define $A(\omega,l)$ to be the event in which, for all $E \in I$, for any
$x,y \in \Lambda_{l} \cap \mathbb{Z}^{d}$ such that $\vert x-y \vert > l_{k}$,
either $\Lambda_{l_{k}}(x)$ or $\Lambda_{l_{k}}(y)$ are $(\gamma, E)$-good.

We shall first use the Borel-Cantelli lemma to show that almost surely $A(\omega,l)$
occurs for all $l$ large enough. Let us define
\begin{eqnarray*}
X_{l} \, := \, \Big\{\omega: A(\omega,l) \, \textrm{is not true at scale} \, l \Big\}.
\end{eqnarray*}
Then we can write
\begin{eqnarray*}
X_{l}&:=&  \Big\{\omega: \exists \, E \in I,\  \exists \, x,y\in \Lambda_{l} \cap \mathbb{Z^{d}}\ \mathrm{with}\
\vert x - y \vert > l_{k}, \\
&&\textrm{such that both} \, \Lambda_{l_{k}}(x)\ \mathrm{and} \  \Lambda_{l_{k}}(y) \, \textrm{are not}
\, (\gamma,E)\textrm{-}\mathrm{good} \Big\}\\
&=& \bigcup_{\substack{x,y \in \Lambda_{l} \cap \mathbb{Z^{d}} \\ \vert x-y \vert > l_{k}}}
\Big\{\omega: \exists \, E \in I, \,\, \textrm{such that both} \, \Lambda_{l_{k}}(x) \ \mathrm{and} \
\Lambda_{l_{k}}(y)
\, \textrm{are not} \, (\gamma,E)\textrm{-}\mathrm{good} \Big\},
\end{eqnarray*}
and by Proposition \ref{multiscale-continuous-case} we obtain
\begin{eqnarray*}
\mathbb{P} (X_{l})  \, \leqslant \, l^{2d} \, (l_{k})^{-2\zeta} \, \leqslant \, l^{-2(\delta\zeta - d)},
\end{eqnarray*}
where the last step follows from (\ref{allowed-interval-for-lk-continuous-case}).  Since $2(\delta\zeta - d)>1$,
it follows that
\begin{eqnarray*}
\sum_{l} \, \mathbb{P} (X_{l}) \, < \, \infty.
\end{eqnarray*}
By the Borel-Cantelli lemma,
almost surely there exists $L(\omega) < \infty$ such that the event $A(\omega,l)$ occurs
for all $l > L(\omega)$.

Since by Lemma \ref{eigenvalues} with probability one, $E_{i}^{\omega,l}$ tends to 0 as $l$ tends to $\infty$,
$E_{i}^{\omega,l} \in I$ for $l$ large enough almost surely. Therefore there exists
$\tilde{\Omega} \subset \Omega$ with $\mathbb{P}(\tilde{\Omega}) =1$
such that for each $\omega \in \tilde{\Omega}$ there is $L_1(\omega) < \infty$
such that for all $l > L_1(\omega)$ and for any
$x,y \in \Lambda_{l} \cap \mathbb{Z}^{d}$ satisfying $\vert x-y \vert > l_{k}$,
either $\Lambda_{l_{k}}(x)$ or $\Lambda_{l_{k}}(y)$ are $(\gamma, E_{i}^{\omega,l})$-good.

Now we take $\omega \in \tilde{\Omega}$ and $l > L_1(\omega)$ and partition the box $\Lambda_{l}(0)$ into
$\Lambda_{l}^{1} := \Lambda_{l-l_{k}}(0)$ and $\Lambda_{l}^{2} := \Lambda_{l}(0) \setminus \Lambda_{l}^{1}$.
We then split up the integral in (\ref{localization-condition-continuous-case}) into the interior
cube $\Lambda_{l}^{1}$  and the corridor $\Lambda_{l}^{2}$
\begin{eqnarray}\label{L1-norm-split-inside-and-corridor-continuous-case}
\int_{\Lambda_{l}} \ud x \, \vert \phi_{i}^{\omega,l}(x) \vert \,\,= \,\,
\int_{\Lambda_{l}^{1}} \ud x \, \vert \phi_{i}^{\omega,l}(x) \vert \,
+ \, \int_{\Lambda_{l}^{2}} \ud x \, \vert \phi_{i}^{\omega,l} (x)\vert\, .
\end{eqnarray}
In the second term, we can use the Schwarz inequality and the fact that the eigenfunctions
are $L^{2}(\Lambda_{l})$-normalized to obtain
\begin{eqnarray}\label{bound-eigenfunction-in-corridor-continous-case}
\int_{\Lambda_{l}^{2}} \ud x \, \vert \phi_{i}^{\omega,l}(x) \vert \, \leqslant \, \vert \Lambda_{l}^{2}\vert^{1/2}
\, \leqslant \, 2^{d} l^{(d-1)/2} \, l_{k}^{1/2} \, \leqslant \, 2^{d} l^{(d - \delta)/2}\,.
\end{eqnarray}
For the first term in (\ref{L1-norm-split-inside-and-corridor-continuous-case}), we shall use the eigenfunction
decay inequality (\ref{eigenfunction-decay-inequality}) of Proposition
\ref{eigenfunction-decay-inequality-continuous-case}.
We cover the \lq\lq interior cube'' $\Lambda_{l}^{1}$ by disjoints subcubes $\Lambda_{j}$ of side $l_{k}/3$.
Let us call $\{x_{j}\}$ their respective centers.
{Then for each $j$ the cube $\Lambda_{l_{k}}(x_{j})$ is included in $\Lambda_{l}$ and
 $\Lambda_{j}$ coincides with $\Lambda_{l_{k}}^{int}(x_{j})$}.\\
Using the Schwarz inequality and Proposition \ref{eigenfunction-decay-inequality-continuous-case},
we obtain for any $j$ the estimate:
\begin{eqnarray*}
\int_{\Lambda_{j}} \ud x \, \vert \phi_{i}^{\omega,l}(x) \vert  &\leqslant& l^{d/2}  \, \Big(\int_{\Lambda_{l}}
\ud x \vert \chi_{l_{k}}^{int}(x_{j})  \phi_{i}^{\omega,l}(x)\vert^{2} \, \Big)^{1/2}\\
&\leqslant& l^{d/2} \,\Big( \kappa \| \chi_{l_{k}}^{int}(x)
(h_{l_{k}}^{\omega}(x_{j}) - E_{i}^{\omega,l})^{-1} \chi_{l_{k}}^{out}(x) \| \Big)^{1/2}.
\end{eqnarray*}
Hence, for any $j$ such that $\Lambda_{j}$ is $(\gamma, E_{i}^{\omega,l})$-good, one has the
following upper bound
\begin{eqnarray}\label{inside-cube-intermediate-estimate-continuous-case}
\int_{\Lambda_{j}} \ud x \, \vert \phi_{i}^{\omega,l}(x) \vert  \, \leqslant \,
l^{d/2} \mathrm{e}^{-\frac{1}{2}\gamma l_{k}}
\, \leqslant \, l^{d/2} \, \mathrm{e}^{-\frac{1}{2}\gamma l^{\delta}}.
\end{eqnarray}

Now, we distinguish two cases.

The first one corresponds to the situation where all cubes $\Lambda_{l_{k}}(x_{j})$ are
$(\gamma,E_{i}^{\omega,l})$-good. It then follows directly from
(\ref{bound-eigenfunction-in-corridor-continous-case}) and
(\ref{inside-cube-intermediate-estimate-continuous-case}) that
\begin{eqnarray}\label{bound1}
l^{-d/2} \, \int_{\Lambda_{l}} \ud x \, \vert \phi_{i}^{\omega,l}(x) \vert  &\leqslant&
2^{d} \frac{l^{(d - \delta)/2}}{l^{d/2}} +
l^{-d/2} \sum_{x_j\in \Lambda_1} \, l^{d/2} \, \mathrm{e}^{-\frac{1}{2}\gamma l^{\delta}}\nonumber\\
& \leqslant& 2^{d} l^{-\delta/2}+ 3^{d} \, \frac{(l-l^{\delta})^{d}}{l^{\delta d}}
\, \mathrm{e}^{-\frac{1}{2}\gamma l^{\delta}}.
\end{eqnarray}

The second case corresponds to the situation when there exists at least one subcube $\Lambda_{l_{k}}(x_{j})$
which is \emph{not} $(\gamma,E_{i}^{\omega,l})$-good. Let us denote by $\tilde{x}$ the center of one
such bad cube. Since $\omega \in \tilde{\Omega}$ and $l > L_1(\omega)$, for
$x,y \in \Lambda_{l} \cap \mathbb{Z}^{d}$
satisfying $\vert x-y \vert > l_{k}$,
either $\Lambda_{l_{k}}(x)$ or $\Lambda_{l_{k}}(y)$ are $(\gamma, E_{i}^{\omega,l})$-good.
It therefore follows that,
outside of a box of side $2l_{k}$ centered at $\tilde{x}$, all other
$\Lambda_{l_{k}}(x_{j})$ are $(\gamma,E_{i}^{\omega,l})$-good. We treat the
good boxes as above, and deal with $\Lambda_{2l_{k}}(\tilde{x})$ by using the 
Schwarz inequality as we did for $\Lambda^2_l$, to obtain:
\begin{eqnarray*}
\int_{\Lambda_{l}^{1}} \ud x \, \vert \phi_{i}^{\omega,l}(x) \vert  &=&
\int_{\Lambda_{l}^{1}\setminus\Lambda_{2l_{k}}(\tilde{x})} \ud x \, \vert \phi_{i}^{\omega,l}(x) \vert
\, + \, \int_{\Lambda_{2l_{k}}(\tilde{x})} \ud x \, \vert \phi_{i}^{\omega,l}(x) \vert\\
&\leqslant& \sum_{x_j\in \Lambda_{l}^{1}\setminus\Lambda_{2l_{k}}(\tilde{x})} \, l^{d/2} \,
\mathrm{e}^{-\frac{1}{2}\gamma l^{\delta}}
+ \, \vert \Lambda_{2l_{k}}(\tilde{x}) \vert^{d/2} \, \\
&\leqslant& l^{d/2} 3^{d} \, \frac{(l-l^{\delta})^{d}}{l^{\delta d}} \,
\mathrm{e}^{-\frac{1}{2}\gamma l^{\delta}} \, + \, (2l)^{d(1-\delta)/2}.
\end{eqnarray*}
From that last bound and from (\ref{bound-eigenfunction-in-corridor-continous-case}), we get
\begin{eqnarray}\label{bound2}
l^{-d/2} \, \int_{\Lambda_{l}} \ud x \, \vert \phi_{i}^{\omega,l}(x) \vert \, \leqslant \,
2^{d} l^{-\delta/2}+ 3^{d} \, \frac{(l-l^{\delta})^{d}}{l^{\delta d}} \, \mathrm{e}^{-\frac{1}{2}\gamma l^{\delta}}
+ \, 2^{d(1-\delta)/2} l^{-d\delta /2}.
\end{eqnarray}
Therefore for any $\omega\in \tilde{\Omega}$ either (\ref{bound1}) or (\ref{bound2}) is
satisfied for all $l$ large enough
and (\ref{localization-condition-continuous-case}) follows.

\hfill $\square$
\subsection{Weak external potentials}\label{Weak external potentials}
In this section we consider a scaled external potential.
Let $v$ be a non-negative, continuous real-valued function defined on the closed unit cube
${\overline \Lambda}_1\subset \mathbb{R}^d$ which satisfies the following two conditions:
\begin{enumerate}
  \item There is a finite, nonempty subset of $\Lambda_{1}$,  $D:=\{y_{j}\}_{j=1}^{n}$ such that
  $v(x) = 0$ if and only if $ x \in D$.
  \item For each $y_{j}\in D$ there are strictly positive numbers $\{\alpha_{j}\}$, $\{c_{j}\}$ such that
\begin{eqnarray}\label{general-definition-weak-external-potential-producing-typeIII-BEC}
 \lim_{x \rightarrow y_{j}} \, \frac{v(x)}{\vert x - y_{j} \vert^{\alpha_{j}}} \, = \, c_{j}.
\end{eqnarray}
\end{enumerate}
We order the $y_{j}$'s in such a way that
$0 < \alpha_{1} \, \leqslant \, \alpha_{2} \, \leqslant \, \dots \, \leqslant \, \alpha_{n}$.

The \textit{one-particle} Schr\"{o}dinger operator with
a \textit{weak} (scaled) external potential in a box $\Lambda_{l}$ is defined by:
\begin{eqnarray}\label{Schrodinger-operator-weak-external-potential}
h_{l} \, = \, -\shalf \Delta_D \, + \, v({x_{1}}/{l},\dots,{x_{d}}/{l}) \ .
\end{eqnarray}
We recall that the eigenfunctions and eigenvalues of $h_{l}$ are denoted by
$\phi_{i}^{l}$ and $E_{i}^{l}$ respectively.
The aim of this section is to prove that our localization condition (\ref{general-localization-definition})
holds for this class of weak potentials.
\begin{lemma}
Let $h_{l}$ be as in (\ref{Schrodinger-operator-weak-external-potential}). Then, for all $i$
\begin{eqnarray}\label{localization-eigenfunctions-weak-external-potential}
\lim_{l \rightarrow \infty} \, \frac{1}{l^{d/2}} \,
\int_{\Lambda_{l}} \, \ud x \, \vert \phi_{i}^{l}(x) \vert \,\, = \,\, 0.
\end{eqnarray}
\end{lemma}

\noindent\textbf{Proof:}
We start by noting that in view of the condition
(\ref{general-definition-weak-external-potential-producing-typeIII-BEC}), for any $\varepsilon > 0$ small enough,
there exists $\delta> 0$ such that for all $j = 1, \dots, n$
\begin{eqnarray}\label{zeros}
(c_{j} - \varepsilon) \vert x - y_{j} \vert^{\alpha_{j}} \, \leqslant \, v(x) \, \leqslant \, (c_{j} + \varepsilon)
\vert x - y_{j} \vert^{\alpha_{j}},
\end{eqnarray}
for all $x\in B(y_{j}, \delta)$, the ball of radius $\delta$ centered at $y_{j}$.
Note also that by continuity there exists a constant $\kappa > 0$ such that $v(x) \geqslant \kappa$,
for all $x \in \Lambda_{1} \setminus \left(\bigcup_{j=1}^{n} B(y_{j}, \delta)\right)$.
We let $K:= \min (\kappa, c_{1} - \varepsilon,\ldots, c_{n} - \varepsilon)$ and
$C:= \max (c_{1} + \varepsilon,\ldots, c_{n} + \varepsilon)$.

The first step in our proof is to obtain an estimate for the eigenvalue $E_{i}^{l}$. To this end, let us denote
by $h_{l}^{(n)}$ the restriction of the Schr\"{o}dinger operator to the region $B(y_{n}, \delta l) $,
with Dirichlet boundary conditions.
Then we have
\begin{eqnarray}\label{weak-potential-localization-DN-bracketing}
h_{l} \, \leqslant \,  h_{l}^{(n)}
\end{eqnarray}
in quadratic form sense (c.f. \cite{RS-IV}, Chapter VIII, Proposition 4). From the inequality (\ref{zeros}),
we obtain
\begin{eqnarray}\label{weak-potential-localization-DN-bracketing-upper-bound}
h_{l}^{(n)} \, \leqslant \, \tilde{h}_{l}^{(n)} \, := \,  \shalf \Delta_{D} \,
+ C \left\vert \frac{x - y_{n}}{l} \right \vert^{\alpha_{n}},
\end{eqnarray}
where the last operator acts on {$L^{2}\big(B(y_{n}, \delta l))$.
Let $U:L^{2} \big(B(y_{n}, \delta l)\big) \mapsto L^{2} \big(B(0, \delta l^{1-\gamma_{n}})\big)$}
be the unitary transformation defined by
\begin{eqnarray}\label{localization-weak-potential-unitary-transormation}
(U \varphi) (x) \, &:=& \, l^{\gamma_{n}/2} \, \varphi(l^{\gamma_{n}} (x - y_{n})),\nonumber
\end{eqnarray}
where $\gamma_{n} := \alpha_{n}/(2 + \alpha_{n})$.
By direct computation, one can check that  $\tilde{h}_{l}^{(n)}=l^{-2\gamma_{n}}\, U\,\hat{h}_{l}^{(n)}\,U^{-1}$
where
\begin{eqnarray*}
\hat{h}_{l}^{(n)}:= (-\shalf \Delta + C \vert x \vert^{\alpha_{n}}),
\end{eqnarray*}
acting on $L^{2} \big(B(0, \delta l^{1-\gamma_{n}})\big)$. Let $0<D_1^{l}\leqslant D_2^{l}\leqslant \ldots$
be the eigenvalues of $\hat{h}_{l}^{(n)}$ and $0<D_1\leqslant
D_2\leqslant \ldots$ the eigenvalues of $\hat{h}^{(n)}$ where
\begin{eqnarray*}
\hat{h}^{(n)}:= (-\shalf \Delta + C \vert x \vert^{\alpha_{n}}),
\end{eqnarray*}
acting on $L^{2} (\mathbb{R}^d)$.
Since for each $i$, $D_i^{l} \to D_i$ as $l\to\infty$, there are constants $\tilde{D}_i$ such that
$D_i^{l} \leqslant  \tilde{D}_{i}$ for all $l$.
Using this and the operator inequalities (\ref{weak-potential-localization-DN-bracketing}) and
(\ref{weak-potential-localization-DN-bracketing-upper-bound}) we finally get
\begin{eqnarray}\label{weak-potential-localization-eigenvalues-upper-bound}
E_{i}^{l} \, \leqslant \, {D}^l_{i} l^{-2\gamma_{n}}\, \leqslant \, \tilde{D}_{i} l^{-2\gamma_{n}}.
\end{eqnarray}

The rest of our proof relies on the methods developed in \cite{McA-Pul}.
We start with some definitions. Let $\Omega^{t}$, for some $t > 0$, to be the set of all continuous
trajectories (paths) $\{\xi(s)\}_{s=0}^{t}$ in ${{\mathbb{R}}}^{d}$ with $\xi(0)=0$, and let $w^{t}$ denote the
normalized Wiener measure on this set. For a given $x \in \mathbb{R}^{d}$,
we define the following characteristic function
\begin{eqnarray*}
\chi_{x,l}(\xi) \, := \, \mathbf{1} \big\{ \xi: \, \xi(s) \in \Lambda_{l}-x, \, \textrm{for all} \,
0 \leqslant s \leqslant t\big\}.
\end{eqnarray*}
We now use the following identity (c.f. \cite{R}),
\begin{eqnarray*}
(\mathrm{e}^{-t h_{l}} \, \phi_{i}^{l})(x) \, = \, \int_{\Omega^{t}} w^{t} (\ud \xi) \,
\mathrm{e}^{\displaystyle{-\int_{0}^{t} \ud s  v((x + \xi(s))/l)}} \, \phi_{i}^{l}(x + \xi(t)) \, \chi_{x,l}(\xi),
\end{eqnarray*}
from which, since $E_{i}^{l}$ is the eigenvalue of $h_{l}$ corresponding to $\phi_{i}^{l}$, we get
\begin{eqnarray}\label{localization-weak-potential-intermediate1}
\vert \phi_{i}^{l}(x) \vert \, \leqslant \, \mathrm{e}^{t E_{i}^{l}} \, \int_{\Omega^{t}} w^{t} (\ud \xi) \,
\mathrm{e}^{\displaystyle{-\int_{0}^{t} \ud s  v((x + \xi(s))/l)}} \,
\vert \phi_{i}^{l}(x + \xi(t)) \vert \, \chi_{x,l}(\xi).
\end{eqnarray}
Now, we insert into the right-hand side of (\ref{localization-weak-potential-intermediate1})
the following bound proved in \cite{D},
\begin{eqnarray*}
\vert \phi_{i}^{l} (x) \vert \, \leqslant c_d  \, (E_{i}^{l})^{d/4},
\end{eqnarray*}
where $c_d:=(e/\pi)^{d/4}$ and we obtain from (\ref{localization-weak-potential-intermediate1}) the following estimate
\begin{eqnarray*}
\vert \phi_{i}^{l}(x) \vert &\leqslant& c_d\, \mathrm{e}^{t E_{i}^{l}} (E_{i}^{l})^{d/4} \,
\int_{\Omega^{t}} w^{t} (\ud \xi) \, \mathrm{e}^{\displaystyle{-\int_{0}^{t} \ud s  v((x + \xi(s))/l)}}
\, \chi_{x,l}(\xi)\\
&=& c_d\, \mathrm{e}^{t E_{i}^{l}} (E_{i}^{l})^{d/4} \,
\int_{\Omega^{t}} w^{t} (\ud \xi) \, \mathrm{e}^{\displaystyle{-\frac{1}{t}\int_{0}^{t} \ud s \, t \, v((x + \xi(s))/l)}}
\, \chi_{x,l}(\xi)\\
&\leqslant& c_d\, \mathrm{e}^{t E_{i}^{l}} (E_{i}^{l})^{d/4} \,
\int_{\Omega^{t}} w^{t} (\ud \xi) \,\frac{1}{t}\int_{0}^{t} \ud s \, \mathrm{e}^{\displaystyle{-t v((x + \xi(s))/l)}}
\, \chi_{x,l}(\xi),
\end{eqnarray*}
where the last step follows from Jensen's inequality. Therefore, integrating over $\Lambda_{l}$ with respect
to $x$, and then changing the order of integration, yields
\begin{eqnarray*}
l^{-d/2} \int_{\Lambda_{l}} \ud x \,\vert \phi_{i}^{l}(x) \vert &\leqslant& \hskip -0.05cm c_d\,
l^{-d/2} \mathrm{e}^{t E_{i}^{l}}
(E_{i}^{l})^{d/4} \int_{\Lambda_{l}} \ud x \, \int_{\Omega^{t}} w^{t} (\ud \xi) \, \frac{1}{t}\int_{0}^{t} \ud s \,
\mathrm{e}^{\displaystyle{-t v((x + \xi(s))/l)}} \,\chi_{x,l}(\xi)\\
&\leqslant& c_d\, l^{-d/2} \mathrm{e}^{t E_{i}^{l}}
(E_{i}^{l})^{d/4} \, \int_{\Omega^{t}} w^{t} (\ud \xi) \, \frac{1}{t}\int_{0}^{t} \ud s
\int_{\{x \in \bigcap_{s'}(\Lambda_{l} - \xi(s'))\}}
\hskip -1cm \ud x \, \mathrm{e}^{\displaystyle{-t v((x + \xi(s))/l)}}.
\end{eqnarray*}
Letting $y = x + \xi(s)$ in the second integral we get
\begin{eqnarray*}
l^{-d/2} \int_{\Lambda_{l}} \ud x \vert \phi_{i}^{l}(x) \vert &\leqslant& c_d\, l^{-d/2} \mathrm{e}^{t E_{i}^{l}}
(E_{i}^{l})^{d/4} \, \int_{\Omega^{t}} w^{t} (\ud \xi) \, \frac{1}{t}\int_{0}^{t} \ud s \\
&.&\int_{\{y - \xi(s) \in \bigcap_{s'}(\Lambda_{l} - \xi(s'))\}} \ud y \, \mathrm{e}^{\displaystyle{-t v(y/l)}}.
\end{eqnarray*}
Since $\bigcap_{s'}(\Lambda_{l} - \xi(s') + \xi(s)) \, \subset \,\Lambda_{l}$ for all $s$, we can now extend the
domain of integration over $y$ to $\Lambda_l$ and use the fact that the Wiener measure $w^t$ is normalized to obtain
\begin{eqnarray}\label{localization-weak-potential-intermediate2}
l^{-d/2} \int_{\Lambda_{l}} \ud x \vert \phi_{i}^{l}(x) \vert &\leqslant& c_d\,\mathrm{e}^{t E_{i}^{l}}
(E_{i}^{l})^{d/4} \, l^{-d/2} \,\frac{1}{t}\int_{0}^{t} \ud s \, \int_{\Lambda_{l}} \ud y \,
\mathrm{e}^{\displaystyle{-t v((y)/l)}}\\
&=& c_d\, \mathrm{e}^{t E_{i}^{l}}
(E_{i}^{l})^{d/4}  \, l^{d/2} \,\int_{\Lambda_{1}} \ud z \,  \mathrm{e}^{-t v(z)}. \nonumber
\end{eqnarray}
Next, we obtain an upper bound for the last integral in (\ref{localization-weak-potential-intermediate2}). We have
\begin{eqnarray}\label{localization-weak-potential-intermediate3}
\int_{\Lambda_{1}} \ud z \,  \mathrm{e}^{-t v(z)} &\leqslant& \sum_{j=1}^{n} \, \int_{B(y_{j}, \delta)} \ud z \,
\mathrm{e}^{-t v(z)} \, \, + \,\,
\int_{\Lambda_{1} \setminus \big(\bigcup_{i=1}^{n} B(y_{j}, \delta)\big)} \ud z \,  \mathrm{e}^{-t v(z)}\\
&\leqslant&  \mathrm{e}^{-t K} \, + \,  \sum_{j=1}^{n} \, \int_{B(y_{j}, \delta)} \ud z \,
\mathrm{e}^{- t K\vert x - y_{j} \vert^{\alpha_{j}}}.
\nonumber
\end{eqnarray}
For each $j$,
\begin{eqnarray*}
\int_{B(y_{j}, \delta)} \ud z \,  \mathrm{e}^{-t K  \vert x - y_{j} \vert^{\alpha_{j}}} \, \leqslant \,
t^{-d/\alpha_{j}} \, K^{d/\alpha_{j}}\,
 \, \int_{\mathbb{R}^{d}} \ud \tilde{z} \, \mathrm{e}^{- \vert \tilde{z} \vert^{\alpha_{j}}} \, \leqslant \,
\tilde{K} \, t^{-d/\alpha_{j}},
\end{eqnarray*}
where $\tilde{K}:=K^{d/\alpha_{1}} \max_j\int_{\mathbb{R}^{d}} \ud \tilde{z} \,  e^{- \vert \tilde{z} \vert^{\alpha_{j}}}$,
which, in view of (\ref{localization-weak-potential-intermediate3}), gives the following bound
\begin{eqnarray*}
\int_{\Lambda_{1}} \ud z \,  \mathrm{e}^{-t v(z)} \, \leqslant \, \mathrm{e}^{-t K} \, + \,  \tilde{K} \sum_{j=1}^{n}
 \, t^{-d/\alpha_{j}}.
\end{eqnarray*}
Now, fixing $t = (E_{i}^{l})^{-1}$, we get from the last inequality and
(\ref{localization-weak-potential-intermediate2})
\begin{eqnarray*}
l^{-d/2} \int_{\Lambda_{l}} \ud x \vert \phi_{i}^{l}(x) \vert \, \leqslant \, c_d\, \mathrm{e}
(E_{i}^{l})^{d/4}  \, l^{d/2} \, \Big(\mathrm{e}^{-K(E_{i}^{l})^{-1}} \, + \, \tilde{K} \sum_{j=1}^{n}
\, (E_{i}^{l})^{d/\alpha_{j}} \Big).
\end{eqnarray*}
Since by (\ref{weak-potential-localization-eigenvalues-upper-bound}), $E_{i}^{l}\to 0$ as $l\to \infty$,
and since we have ordered the $\alpha_{i}$'s such that
$\alpha_{1} < \alpha_{2} < \dots < \alpha_{n}$, there exist new constants $A_{i}$ such that the following
bound holds for $l$ large enough
\begin{eqnarray}\label{localization-weak-potential-intermediate4}
l^{-d/2} \int_{\Lambda_{l}} \ud x \vert \phi_{i}^{l}(x) \vert \, \leqslant \, A_{i} \,
l^{d/2} \big(E_{i}^{l}\big)^{d(1/4 + 1/\alpha_{n})} \, = \, A_{i} \,
l^{d/2} \big(E_{i}^{l}\big)^{d(2-\gamma_{n})/(4\gamma_{n})}.
\end{eqnarray}
Inserting the bound (\ref{weak-potential-localization-eigenvalues-upper-bound}),
we finally obtain for $l$ large enough
\begin{eqnarray*}
l^{-d/2} \int_{\Lambda_{l}} \ud x \vert \phi_{i}^{l}(x) \vert \, \leqslant A_{i}
\tilde{D}_i^{d(2-\gamma_{n})/(4\gamma_{n})} \, l^{-d(1-\gamma_n)/2}
\end{eqnarray*}
and the lemma follows since $\gamma_n<1$. \hfill $\square$

\section*{Acknowledgments}

One of the authors (Th.Jaeck) is supported by funding from the UCD Ad Astra Research Scholarship.


\end{document}